# Superhumps and flickering in V1316 Cygni

David Boyd, Christopher Lloyd, Robert Koff, Thomas Krajci, Bart Staels, Jerrold Foote, William Goff, Tonny Vanmunster, Lewis Cook, Joseph Patterson

**Abstract**

We present analysis and results of a coordinated CCD photometry campaign to observe the 2006 June superoutburst of the cataclysmic variable V1316 Cyg involving 8 longitudinally-distributed observers. The outburst peaked at magnitude 15.03 on June 10, declined at a rate of 0.14 mag day$^{-1}$, lasted 11 days and had an amplitude above quiescence of 2.4 magnitudes. We detected common superhumps for the first time, thereby confirming that V1316 Cyg is a member of the UGSU class of dwarf novae. We observed a transition to late superhumps two-thirds of the way through the outburst with an associated phase shift of 0.50 +/- 0.06 cycles. The mean common superhump period before this transition was 0.07685 +/- 0.00003 d and the mean late superhump period following the transition was 0.07654 +/- 0.00002 d. The common superhump period decreased at a rate dP/dt = –5.1 +/- 1.7 *10$^{-5}$ cycle$^{-1}$. At the onset of late superhumps, there was a transient shift in power from the superhump fundamental frequency to its first harmonic and back again. We detected an orbital period of 0.0740 +/- 0.0002 d giving a fractional superhump period excess of 0.038 +/- 0.003 and a mass ratio of 0.167 +/- 0.010. A scalegram analysis of the flickering behaviour of V1316 Cyg found that the α and Σ parameters characterising flickering changed significantly during the superoutburst. We also found flickering to be at a relatively much lower level at the beginning of the superoutburst and during two "normal" outbursts.

**History**

V1316 Cyg was discovered by Romano [1], originally labelled GR141, and classified as LP, presumably meaning a long period variable. Romano [2] later reclassified it as a dwarf nova with designation V1316 Cyg. A spectrum obtained during an outburst by Bruch & Schimpke [3] showed Balmer emission lines on a strong blue continuum, confirming it was a dwarf nova.

V1316 Cyg has acquired a UGSU classification but it is unclear how. Downes & Shara [4] cite Bruch, Fischer & Wilmsen [5] as the type reference but the latter appear not to have classified it, and indeed not to have observed it in outburst. So far, this classification seems to have been unsupported by observational evidence.

Correct identification of V1316 Cyg has also been a problem. In figure 19 of Bruch, Fischer & Wilmsen [5], V1316 Cyg was incorrectly identified as the nearby magnitude 14.8 star. In Downes & Shara [4] this mis-identification was continued. This was finally resolved only in 2000 September when V1316 Cyg was correctly identified as the faint star 12 arcsec to the east of the magnitude 14.8 star [6]. The situation was further complicated by the detection of low amplitude variability in the magnitude 14.8 star in the years before 2000. Downes et al. [7] give a correct identification chart for V1316 Cyg and list the coordinates as RA 20h 12m 13.62s, Dec +42° 45' 51.5" (J2000).

These difficulties mean that the reliable historical record of observations of V1316 Cyg only begins in 2000. The AAVSO International Variable Star Database [8] contains one well-documented outburst which was first reported by Simonsen on 2003 September 21 and lasted about 17 days. During this outburst, estimates were contributed by 11 observers, 6 observing visually and 5 using a CCD. Only one brief time-series photometry run was obtained which did not confirm the presence of superhumps. The maximum reliable magnitude reported during this outburst was 15.2. A search of the BAA Variable Star Section (BAAVSS) database did not reveal any additional observations.

**Detection of short, low amplitude outbursts**

As part of the BAAVSS Recurrent Objects Programme [9], V1316 Cyg has been monitored regularly. During 2005 several unusually short outbursts of V1316 Cyg were detected [10]. These were of lower amplitude than the outburst reported in 2003, reaching about magnitude 16. They typically lasted 1-2 days and recurred sometimes after only 10 days. Time-series photometry during one of these outbursts on 2005 September 5 at magnitude 16.0 did not detect any sign of superhumps. Whether these are the "normal" outbursts of V1316 Cyg is unclear. Further observations are needed to understand them better.

**Photometric observations during 2006 June superoutburst**

The quiescent magnitude of V1316 Cyg is around 17.4 but varies continuously by several tenths of a magnitude due to flickering. Following a period at quiescence, V1316 Cyg was observed brighter and rising on 2006 June 7 and on June 9 superhumps were detected in the light curve for the first time. This finally confirmed the postulated UGSU classification of V1316 Cyg. Figure 1 shows the field of V1316 Cyg on June 9. Henceforth we will refer to dates in the truncated form JD = JD - 2,453,000.

The Centre for Backyard Astrophysics [11] is a global network of small telescopes set up to study the periodic behaviour of cataclysmic variables. Following announcement of the outburst of V1316 Cyg, CBA observers began an intensive observing campaign. 8 observers were involved, 3 in Europe and 5 spread across America, collecting between them 32 time-series photometry runs comprising 123 hours of photon collection and over 5000 magnitude measurements on 18 out of 24 nights. This comprehensive coverage of the outburst continued until the star had returned to quiescence and is an impressive example of the power of a longitudinally-distributed network of observers who can be quickly mobilised and have ready access to capable equipment.

CCD images were calibrated and reduced by each observer. All images were dark-subtracted and flat-fielded then measured using differential aperture photometry. Comparison stars from the AAVSO f-chart for V1316 Cyg (dated 020301) were used with magnitudes from photometry by Henden given in Sumner [12]. Close proximity of the star labelled 148 on the AAVSO f-chart necessitated care to ensure that light from that star did not contaminate either the photometric aperture for V1316 Cyg or measurements of sky background.

The observations obtained are listed in Table 1 and the instrumentation used in Table 2. In almost every case, observers worked unfiltered to maximise signal to noise ratio as the principal objective was to record the time-varying behaviour of the light curve as accurately as possible. Given the different spectral responses of the cameras used, some variation between the magnitudes measured by different observers is inevitable. The discrepancy between concurrent measurements was generally less than 0.1 magnitude.

Figure 2 shows the light curve of the superoutburst. The outburst started on JD 894, or possibly the previous day, and lasted 11 days reaching a maximum magnitude of 15.03 on JD 897, an outburst amplitude of 2.4 magnitudes. Maximum light was followed by a steady decline at a rate of 0.14 mag day$^{-1}$ for 8 days before a rapid drop at the end of the outburst on JD 905. Immediately following the main outburst, there was a brief rebrightening from 17.1 to 16.5 before the star returned to its previous quiescent level below magnitude 17. Such rebrightenings are not uncommon in UGSU stars. It was similar to one of the short, low amplitude outbursts noted earlier. The coordinates of V1316 Cyg measured from images during the outburst agree with those given in Downes et al. [7] to better than 0.2 arcsec.

**Analysis of superhumps**

13 of the 32 observing runs obtained during the campaign were longer than 4 hours and light curves of these are shown in Figure 3. All the light curves are drawn at the same time and magnitude scale. The presence of superhumps is a prominent feature of the early runs with their amplitude gradually diminishing from 0.5 magnitude at the peak of the outburst to around 0.1 magnitude as the outburst ended. The structure of the light curves also becomes more complex towards the end of the outburst.

Times of maximum of the superhumps which could be resolved in the individual light curves were determined using a quadratic fit. For three of the later runs, in which the superhumps were less clearly defined, we subtracted the linear trend then synchronously summed the data on the period 0.07654 d before fitting a maximum. In all, 40 maxima were measured of which 29 are shown in Figure 3. A preliminary analysis showed that the superhump phase remained relatively stable up to the end of JD 901. The times of maximum from JD 896 to 901 inclusive were therefore used to determine a preliminary superhump maximum ephemeris of
$$HJD\ 2{,}453{,}896.49647 + 0.07685 * E.$$

Using this ephemeris, times of maximum throughout the outburst were computed and used to assign superhump cycle numbers and calculate O-C values for each observed maximum. These are listed in Table 3 and plotted in Figure 4. The O-C diagram clearly distinguishes between a "common superhump" regime up to at least cycle 70 (JD 901), at which point a transition took place (during JD 902) to a "late superhump" regime (van der Woerd et al. [13]) which was in place by cycle 90 (JD 903). This transition involved a phase shift of 0.50 +/- 0.06 cycles.

From linear fits to the times of maximum in the common and late superhump regions, we found the mean common superhump period to be 0.07685 +/- 0.00003 d and the mean late superhump period to be 0.07654 +/- 0.00002 d. We tested both linear and quadratic fits in the common superhump region. The improvement in chi-squared of the quadratic over the linear fit was significant at the 2% level. The quadratic fit gave a period rate of change dP/dt = -5.1 +/- 1.7 *10$^{-5}$ cycle$^{-1}$ indicating a slowly decreasing period over this interval.

**Period analysis**

To ensure stability when subtracting the mean and trend from each time-series, we excluded time-series shorter than 0.08 days from the period analysis. The remaining 25 observing runs had their means and linear trends subtracted and were then combined. To analyse the frequency content of this combined light curve, we separated it into three segments: (a) before transition (JD 896-901 inclusive), (b) from the onset of transition to the end of the outburst (JD 902-905 inclusive), (c) after the outburst (JD 910-917 inclusive - we had no data for JD 906-909). These segments are shown in Figure 5. Period analysis using a data compensated discrete Fourier transform was carried out on each segment separately with the CLEANest algorithm in the PERANSO software [14]. Power spectra of these analyses are shown in Figure 6 and the frequencies containing most power in each spectrum are listed in Table 4.

(a) In the early and middle part of the outburst up to the onset of phase transition (JD 896-901) there is a strong superhump signal at 13.00 +/- 0.05 cycles day$^{-1}$. Removing this superhump signal leaves small signals at 11.95, 12.86 and 13.45 cycles day$^{-1}$. There is a weak signal at the first harmonic of the superhump frequency, which is probably due to the non-sinusoidal shape of the light curve.

(b) Between the onset of phase transition and the end of the outburst (JD 902-905), power in the superhump signal appears to be divided between its fundamental frequency at 12.96 +/- 0.07 cycles day$^{-1}$ and first harmonic at 26.07 +/- 0.07 cycles day$^{-1}$, while a strong signal at 13.54 +/- 0.14 cycles day$^{-1}$ appears. Removing this latter signal leaves the superhump and its first harmonic essentially unchanged and the reverse is also true so they appear to be unconnected.

(c) Beyond the end of the outburst (JD 910-917), our coverage is relatively poor so discrimination against alias signals in the power spectrum is weak. The strongest signal is the superhump frequency at 13.04 +/- 0.04 cycles day$^{-1}$. Removing this frequency and its aliases leaves little remaining power.

To examine what is happening around the interesting phase transition region in more detail, we analysed the light curves for JD 902, 903, 904 and 905 separately. The power spectra for these four days are shown in Figure 7 and the prominent frequencies listed in Table 5. Over this interval, power appears to shift from the superhump fundamental frequency into its first harmonic and then back again.

In the interval JD 896-901, we find a mean common superhump period of 0.07693 +/- 0.00027 d with mean peak-to-peak amplitude 0.23 mag. In the interval JD 903-917, the mean late superhump period is 0.07654 +/- 0.00010 d and mean peak-to-peak amplitude 0.10 mag. These periods are consistent with the results obtained from linear fits to the times of superhump maximum. Averaged phase diagrams for common and late superhumps are given in Figure 8. The phase diagram for late superhumps shows a double peak as expected from the presence of a strong first harmonic signal immediately after the transition. A weak superhump signal is present for at least 12 days (160 cycles) after the end of the outburst.

Based on available knowledge of UGSU cataclysmic variables (see for example Patterson et al. [15] and Pearson [16]) and assuming V1316 Cyg is not an abnormal member of its class, we estimate the fractional superhump period excess is likely to be ~0.04 and therefore the orbital period to be ~0.074 d (frequency 13.5 cycles day$^{-1}$). We note that the signals at 13.54 cycles day$^{-1}$ for JD 902-905 and 13.48 cycles day$^{-1}$ for JD 910-917 listed in Table 4, plus the weak residual signal at 13.45 cycles day$^{-1}$ for JD 986-901 noted in (a) above, are close to this frequency. Analysing all the data together yields a small but persistent periodic signal at 0.0740 +/- 0.0002 d (frequency 13.52 +/- 0.03 cycles day$^{-1}$) with peak-to-peak amplitude 0.05 mag which we interpret as the orbital period. Radial velocity measurements or further time-series photometry at quiescence are required to confirm this orbital period. The fractional common superhump period excess $\varepsilon = (P_{sh} - P_{orb}) / P_{orb}$, where $P_{sh}$ is the common superhump period and $P_{orb}$ is the orbital period, is then 0.038 +/- 0.003. Using the relationship $\varepsilon = 0.18q + 0.29q^2$ (Patterson et al. [15]) gives a mass ratio q = 0.167 +/- 0.010.

The period analysis was repeated using the ANOVA method (Schwarzenberg-Czerny [17]) which gave results fully consistent with the above analysis. All the main features noted above were present in both solutions.

What physical explanations can we invoke for these observations? During the early and middle part of the outburst, from JD 896 to JD 901, we see a strong common superhump signal caused by tidal and thermal stresses induced by the secondary as it passes the maximum radius of an eccentric precessing accretion disc (Hellier [18]). The slow reduction in the period may be explained by the accretion disc emptying and shrinking. During JD 902 we see transition to a late superhump regime where the dominant light source is now the hot spot where the accretion stream impacts the disc (Rolfe et al. [19]). Maximum light is produced at the point when the accretion stream generates most energy as it impacts the disc. This occurs where the eccentric disc has minimum radius, thus explaining the 0.5 cycle superhump phase change. We note that in V1316 Cyg this transition from common to late superhumps occurs at a point approximately two-thirds of the way through the outburst rather than at the end of the outburst as in IY UMa (Patterson et al. [20]) or well after the outburst as in 1RXS J232953.9+062814 (Skillman et al. [21]). The temporary increase in power at the first harmonic of the superhump frequency during JD 903 and 904 may possibly indicate a brief resurrection of the common superhump mechanism which then dies away again by JD 905.

**Analysis of flickering**

In common with many other dwarf novae, V1316 Cyg exhibits flickering. This is a stochastic variation in the light output of the system covering many timescales which is thought to be caused by irregularities in the flow of material, either in the accretion stream or at the inner edge of the disc. Flickering is most apparent in quiescence and is primarily responsible for the variation in the observed quiescent magnitude.

A well-established technique for analysing flickering behaviour is the scalegram (Fritz & Bruch [22]) which is a log-log plot of time scale against normalised amplitude. It is obtained by a wavelet analysis of the light curve using a base function which is designed to represent well the characteristic transient sharply peaked behaviour of flickering in the light curve. This complements Fourier analysis which is optimised for periodic sinusoidal variation. Analysis of the light curve of an object creates a track in the scalegram. The two important parameters of a track are α, the slope of the linear segment of the track which is a measure of the variation of flickering power with time scale, and Σ, the height of the track at a reference time scale which indicates the overall strength of flickering. A positive value of α means there is more flickering power at longer time scales. The flickering behaviour of an object can therefore be represented by its position in α-Σ space. It appears empirically that cataclysmic variables of the same type tend to cluster together in α-Σ space, as shown in figures 10-15 in Fritz & Bruch [22]. This potentially offers assistance in classifying variables of unknown type based on their flickering behaviour.

The scalegram in Figure 9 shows results of a wavelet analysis of the JD 896 (June 9) data of V1316 Cyg and a 15.1 magnitude comparison star. The results were not sensitive to which base or mother wavelet was used so wavelet C6 was adopted for the analysis. Random noise has no preferred time scale so the scalegram of a constant star is essentially flat in this diagram. At the shortest time scales the scalegram of a flickering source will be increasingly dominated by noise and the flickering spectrum will flatten out to the level of the noise in the data. If the noise level can be determined then this can be subtracted from the values at longer time scales to provide a more reliable spectrum (see Fritz & Bruch [22]). All the scalegrams generated here show some indication of flattening so it has been assumed that the value of *log(s)* at the shortest timescale is entirely due to noise and this value has been subtracted from the others to produce the noise-corrected spectrum. Generally the noise correction makes a significant difference to only one or two of the weakest points in the spectrum. At the longest time scales the scalegram is under sampled and the values are unreliable. These points correspond to the time scale of the data length or longer and can be safely ignored. The parameters of the flickering spectrum are determined from the shortest possible time scale up to something less than the orbital period, typically 60 minutes, and all the runs used were substantially longer than this.

Scalegrams have been constructed for all the data sets and 12 of these which provided reliable values of $\alpha$ and $\Sigma$ are shown in Figure 10. The uncertainties are typically 0.2 in both parameters. All these values come from the slow decline in magnitude following the maximum. Generally the later runs are fainter and so have poorer signal to noise, also they tend to have longer exposure times and so the shorter time scales of the scalegrams are less well sampled. Weighted mean values of $\alpha$ and $\Sigma$ have been calculated for each day as plotted in Figure 11, which shows their chronological development as the outburst declined from maximum. In the $\alpha$-$\Sigma$ plane the scalegrams tend to cluster in two groups corresponding to the periods JD 896-899 (June 9-12) and JD 900-904 (June 13-17) although there is some overlap. The earlier points have $\alpha \sim 2.7$ and $\Sigma \sim -2.7$, which indicate an unusually steep gradient but a rather weak flickering level for the start of the superoutburst. These values tend to come from the better sampled scalegrams so should be reliable. The later values of $\alpha \sim 2.0$ and $\Sigma \sim -2.4$ are more typical of UGSU-type dwarf novae in superoutburst (see Figure 15 of Fritz & Bruch [22]). This transition in the $\alpha$-$\Sigma$ plane occurs about two days before the large phase change in the superhump cycle but it is not clear what association there may be, if any, between these events. Despite the relatively large uncertainties in the individual $\alpha$ and $\Sigma$ parameters the correlation between them is clear. The Spearman Rank Correlation is -0.62 with a probability of 3% that this is due to chance. However, the negative correlation is very unusual as only 2 systems (TT Ari and RR Pic) of the 19 studied by Fritz & Bruch [22] behave in this way.

As well as the positive observation of flickering there are two interesting negative observations to report. On the night the outburst was discovered and before superhumps had developed, JD 894 (June 7), there was no significant flickering, but the signal to noise of the observations was sufficient to see any of the subsequent flickering reported here. The implication of this is that the usual flickering mechanism was either switched off or swamped immediately before the superhumps became visible. The data after JD 904 unfortunately have a high noise level so it is not possible to follow the behaviour of the flickering into quiescence. However, in data obtained on JD 1003 (September 24), when the system was undergoing one of its short, low amplitude outbursts at magnitude 16.1, there was also no significant flickering. Observations reported by Shears Boyd & Poyner [10] of another of V1316 Cyg's many brief outbursts on JD 619, a year prior to the superoutburst, show a suggestion of some activity but this is substantially less than the flickering seen during the superoutburst (see Figure 12). The mechanism generating the flickering in V1316 Cyg appears to be quite fragile; it can switch on and off on a short time scale and also evolves quickly. For most CV's flickering is most visible at quiescence, but in this system that regime has not yet been thoroughly explored.

**Conclusions**

A coordinated CCD photometry campaign involving 8 longitudinally-distributed observers has obtained comprehensive coverage of the 2006 June superoutburst of V1316 Cyg. The outburst peaked at magnitude 15.03 on June 10 and subsequently declined at a rate of 0.14 mag day$^{-1}$. It lasted 11 days and had an amplitude above quiescence of 2.4 magnitudes. Our detection of common superhumps confirms for the first time that V1316 Cyg is a member of the UGSU class of dwarf novae. We observed a transition to late superhumps two-thirds way through the outburst with an associated phase shift of 0.50 +/- 0.06 cycles. This contrasts with other systems in which this transition has occurred either at the end of the superoutburst or later. We measured the mean common superhump period before this transition as 0.07685 +/- 0.00003 d and the mean late superhump period following the transition as 0.07654 +/- 0.00002 d. We found the common superhump period slowly decreased at a rate of $dP/dt = -5.1$ +/- $1.7 *10^{-5}$ cycle$^{-1}$. At the onset of late superhumps, we observed a transient shift in power from the superhump fundamental frequency to its first harmonic and back again, possibly indicating a temporary re-growth of common superhumps. We detected a small orbital signal with period 0.0740 +/- 0.0002 d giving a fractional common superhump period excess of 0.038 +/- 0.003. Using the relationship between superhump period excess and mass ratio published in [15] gives a corresponding mass ratio of 0.167 +/- 0.010. From a scalegram analysis of the flickering behaviour of V1316 Cyg, we found that the $\alpha$ and $\Sigma$ parameters characterising flickering changed significantly during the outburst, evolving towards values similar to those observed in other UGSU-type dwarf novae in superoutburst. At the beginning of the outburst, before superhumps formed, flickering appeared to be absent. A similar analysis of flickering during two short, low amplitude outbursts of V1316 Cyg, respectively one year before and three months after this superoutburst, found flickering to be at a substantially lower level than during the superoutburst.

**Acknowledgements**

We acknowledge with thanks variable star observations from the AAVSO International Database contributed by observers worldwide and used in this research. We thank Clive Beech for his assistance in extracting information from the BAAVSS data archives. We are also grateful for the constructive comments of the referee.

**Addresses**

- DB: BAA Variable Star Section & CBA Oxford, 5 Silver Lane, West Challow, Wantage, Oxon, OX12 9TX, UK [drsboyd@dsl.pipex.com]
- CL: Department of Physics and Astronomy, Open University, Milton Keynes, MK7 6AA, UK [c.lloyd@open.ac.uk]
- RK: CBA Colorado, 980 Antelope Drive West, Bennett, CO 80102, USA [bob@antelopehillsobservatory.org]

- TK: CBA New Mexico, PO Box 1351 Cloudcroft, New Mexico 88317, USA [tom_krajci@tularosa.net]
- BS: CBA Flanders (Alan Guth Observatory), Koningshofbaan 51, B-9308 Hofstade, Belgium [staels.bart.bvba@pandora.be]
- JF: CBA Utah, 4175 East Red Cliffs Drive, Kanah, UT 84741, USA [jfoote@scopecraft.com]
- WG: 13508 Monitor Ln., Sutter Creek, CA 95685, USA [b-goff@sbcglobal.net]
- TV: CBA Belgium, Walhostraat 1A, B-3401 Landen, Belgium [tonny.vanmunster@cbabelgium.com]
- LC: CBA Concord, 1730 Helix Court, Concord, CA 94518, USA [lew.cook@gmail.com]
- JP: Department of Astronomy, Columbia University, 550 West 120[th] Street, New York, NY 10027, USA [jop@astro.columbia.edu]

| Date (UT) 2006 | Start time (HJD) 2,453,000+ | Duration (h) | No of images | Exposure (s) | Filter | Mean mag | Observer |
|---|---|---|---|---|---|---|---|
| June 3 | 890.41855 | 0.20 | 10 | 60 | C | 17.17 | DB |
| June 4 | 891.44024 | 0.07 | 3 | 60 | C | 17.61 | DB |
| June 5 | 892.41353 | 0.11 | 6 | 60 | C | 17.60 | DB |
| June 7 | 894.42286 | 3.08 | 132 | 60 | C | 15.82 | DB |
| June 9 | 896.41491 | 4.40 | 303 | 60 | C | 15.21 | DB |
| June 10 | 897.41802 | 2.32 | 132 | 40 | C | 15.03 | DB |
| June 11 | 897.67339 | 6.50 | 349 | 60 | C | 15.10 | TK |
| June 11 | 898.38260 | 5.16 | 345 | 50 | C | 15.09 | BS |
| June 11 | 898.41097 | 0.66 | 53 | 30 | C | 15.10 | DB |
| June 12 | 898.67204 | 7.24 | 380 | 60 | C | 15.18 | TK |
| June 12 | 898.72601 | 5.60 | 213 | 90 | C | 15.19 | JF |
| June 12 | 899.39276 | 4.84 | 194 | 80 | C | 15.39 | TV |
| June 12 | 899.42837 | 2.48 | 103 | 40 | C | 15.33 | DB |
| June 13 | 899.72445 | 5.18 | 200 | 90 | C | 15.39 | JF |
| June 14 | 900.71855 | 5.18 | 110 | 120 | C | 15.57 | RK |
| June 15 | 901.66728 | 6.01 | 131 | 120 | C | 15.65 | RK |
| June 15 | 902.43517 | 3.85 | 205 | 40 | C | 15.80 | DB |
| June 16 | 902.71841 | 5.80 | 220 | 90 | C | 15.73 | JF |
| June 16 | 902.82319 | 2.33 | 67 | 75 | C | 15.72 | LC |
| June 16 | 903.39350 | 4.74 | 268 | 60 | C | 15.81 | BS |
| June 17 | 903.69402 | 5.67 | 102 | 120 | C | 15.99 | RK |
| June 17 | 903.76731 | 3.76 | 306 | 30 | C | 16.04 | WG |
| June 17 | 903.80746 | 3.09 | 128 | 45 | C | 15.86 | LC |
| June 17 | 904.43174 | 1.12 | 34 | 50 | C | 16.05 | DB |
| June 17 | 904.43509 | 3.84 | 142 | 80 | C | 15.93 | TV |
| June 17 | 904.45889 | 1.31 | 75 | 60 | C | 16.04 | BS |
| June 18 | 904.66404 | 6.32 | 151 | 120 | C | 16.19 | RK |
| June 19 | 905.80412 | 2.89 | 134 | 60 | C | 16.87 | WG |
| June 21 | 908.42784 | 0.27 | 17 | 60 | C | 17.13 | DB |
| June 23 | 909.84538 | 1.83 | 46 | 120 | V | 16.46 | WG |
| June 24 | 910.65786 | 6.62 | 161 | 120 | C | 16.76 | RK |
| June 26 | 912.65752 | 6.61 | 163 | 120 | C | 17.63 | RK |
| June 27 | 914.49703 | 0.32 | 21 | 60 | C | 17.65 | DB |
| June 28 | 915.43732 | 1.35 | 48 | 60 | C | 17.65 | DB |
| June 30 | 917.43604 | 2.77 | 151 | 60 | C | 17.40 | DB |

Table 1: Log of observations.

| Observer | Instrumentation |
|---|---|
| DB | 0.35-m f/5.3 SCT + SXV-H9 CCD camera |
| LC | 0.73-m reflector + HSV-H9 CCD camera |
| JF | 0.60-m f/3.4 reflector + ST-8e CCD camera |
| WG | 0.40-m Newtonian + ST-8 CCD camera |
| RK | 0.25-m f/10 SCT + AP47 CCD camera |
| TK | 0.28-m f/10 SCT + ST-7E CCD camera |
| BS | 0.28-m f/6.3 SCT + MX716 CCD camera |
| TV | 0.35-m f/6.3 SCT + ST-7XME CCD camera |

Table 2: Instrumentation used.

| Cycle no | Time of maximum (HJD) 2,453,000+ | O-C (cycles) |
|---|---|---|
| 0 | 896.49364 | -0.0369 |
| 1 | 896.57119 | -0.0277 |
| 13 | 897.49622 | 0.0099 |
| 16 | 897.72912 | 0.0408 |
| 17 | 897.80442 | 0.0206 |
| 18 | 897.88181 | 0.0277 |
| 25 | 898.41436 | -0.0421 |
| 26 | 898.49084 | -0.0468 |
| 27 | 898.56732 | -0.0516 |
| 29 | 898.72717 | 0.0285 |
| 30 | 898.80147 | -0.0045 |
| 30 | 898.80178 | -0.0005 |
| 31 | 898.88115 | 0.0323 |
| 31 | 898.88054 | 0.0244 |
| 32 | 898.95674 | 0.0160 |
| 38 | 899.41656 | -0.0002 |
| 39 | 899.49262 | -0.0105 |
| 40 | 899.57052 | 0.0033 |
| 43 | 899.80299 | 0.0284 |
| 44 | 899.87851 | 0.0112 |
| 56 | 900.80212 | 0.0304 |
| 57 | 900.87940 | 0.0361 |
| 68 | 901.71608 | -0.0760 |
| 69 | 901.79846 | -0.0040 |
| 70 | 901.87491 | -0.0091 |
| 81 | 902.73635 | 0.2010 |
| 83 | 902.88946 | 0.1935 |
| 83 | 902.88070 | 0.0794 |
| 94 | 903.74889 | 0.3775 |
| 95 | 903.82769 | 0.4029 |
| 96 | 903.90185 | 0.3680 |
| 96 | 903.90337 | 0.3877 |
| 96 | 903.90149 | 0.3633 |
| 104 | 904.51693 | 0.3721 |
| 109 | 904.90046 | 0.3631 |
| 121 | 905.81943 | 0.3219 |
| 122 | 905.89548 | 0.3115 |
| 185 | 910.70928 | -0.0453 |
| 211 | 912.70490 | -0.0758 |
| 273 | 917.45381 | -0.2771 |

Table 3: Observed superhump maxima.

| JD 896 - 901 | | JD 902 - 905 | | JD 910 - 917 | |
|---|---|---|---|---|---|
| Freq c d$^{-1}$ | Power | Freq c d$^{-1}$ | Power | Freq c d$^{-1}$ | Power |
| 13.00 | 651 | 13.54 | 96 | 13.04 | 45 |
| 12.04 | 266 | 26.07 | 90 | 13.48 | 40 |
| 13.91 | 202 | 12.96 | 86 | 12.61 | 38 |
| 15.09 | 162 | 12.03 | 69 | 11.99 | 34 |
| 10.00 | 150 | 14.54 | 62 | 14.10 | 32 |

Table 4: Frequencies containing most power in the spectra in Figure 6.

| JD 902 | | JD 903 | | JD 904 | | JD 905 | |
| --- | --- | --- | --- | --- | --- | --- | --- |
| Freq c d$^{-1}$ | Power | Freq c d$^{-1}$ | Power | Freq c d$^{-1}$ | Power | Freq c d$^{-1}$ | Power |
| 13.18 | 78 | 25.93 | 45 | 26.29 | 30 | 13.18 | 48 |
| 15.90 | 43 | 11.82 | 42 | 13.47 | 26 | 26.93 | 9 |
| 5.23 | 21 | 14.11 | 34 | 10.60 | 24 | | |

Table 5: Frequencies containing most power in the spectra in Figure 7.

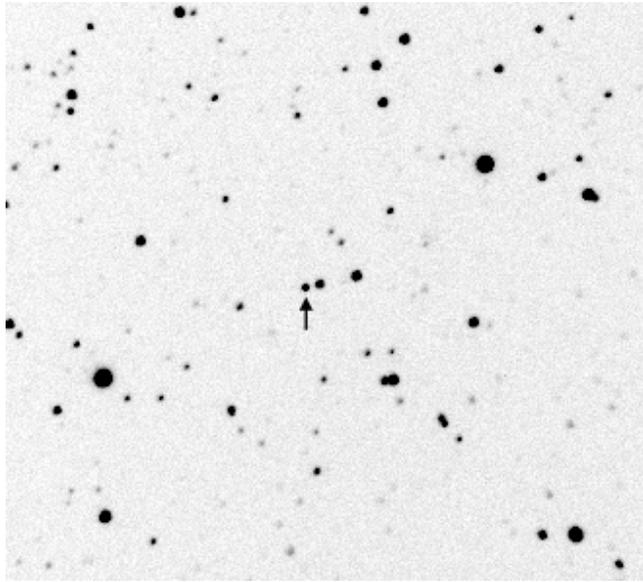

Figure 1: V1316 Cyg on 2006 June 9, field 8' x 8', north at top.

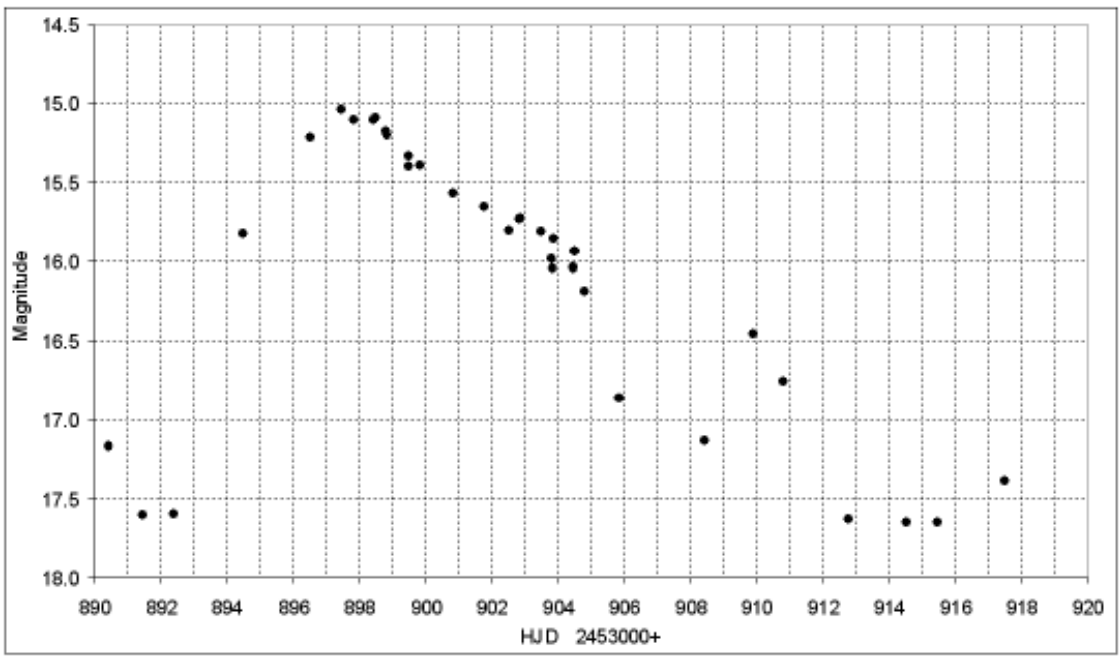

Figure 2: Light curve of 2006 June superoutburst, mean magnitudes for each run.

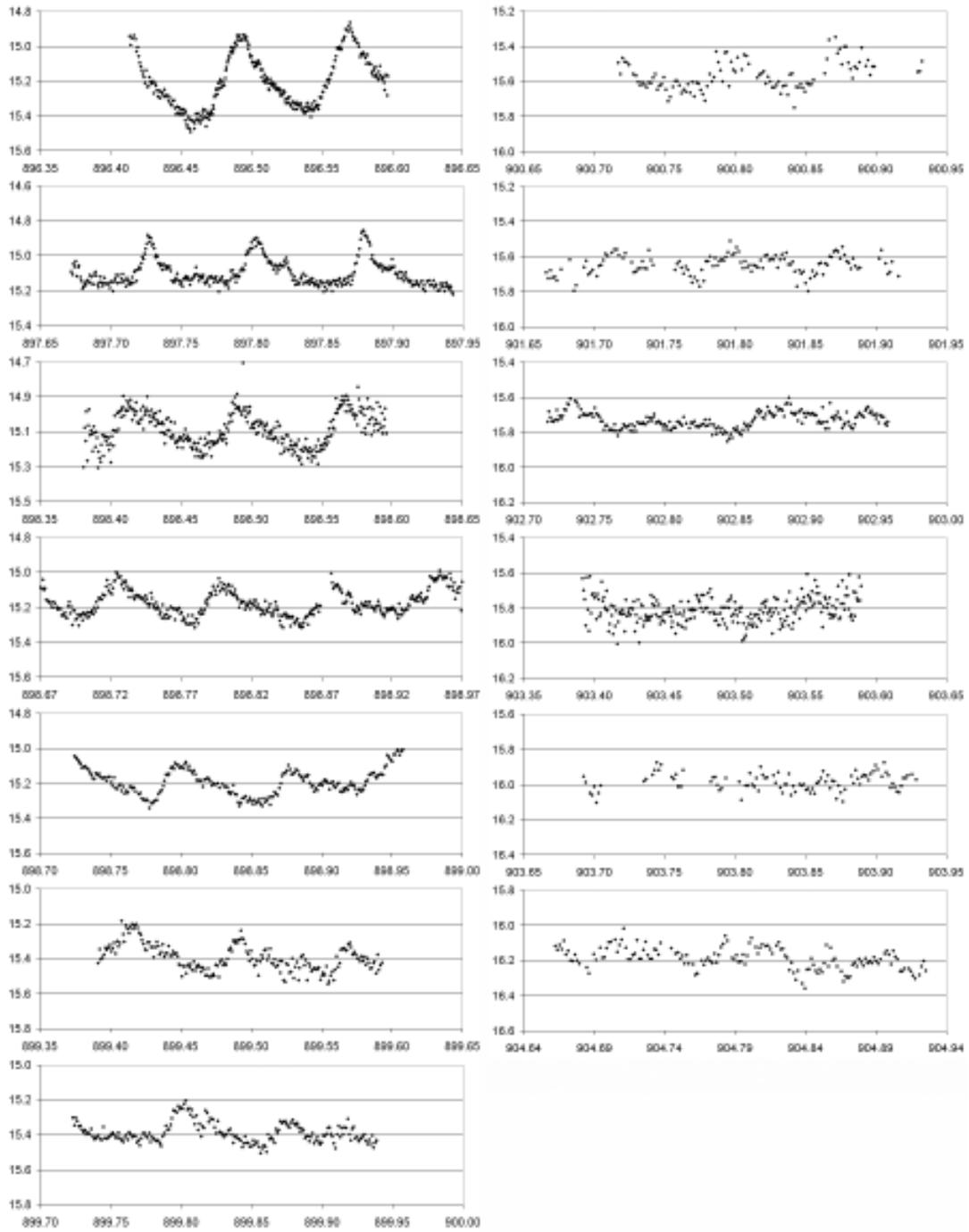

Figure 3: Light curves longer than 4 hours during the outburst. Horizontal axes are JD-2453000. Vertical axes are unfiltered magnitudes. All plots have the same time and magnitude scale.

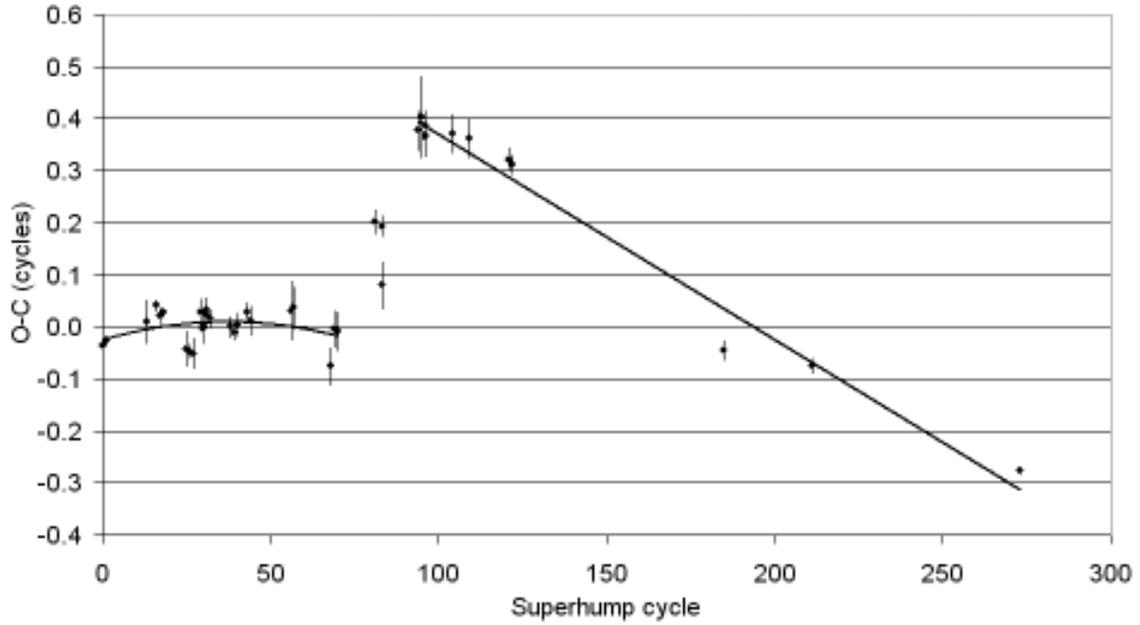

Figure 4: O-C diagram for superhump maxima.

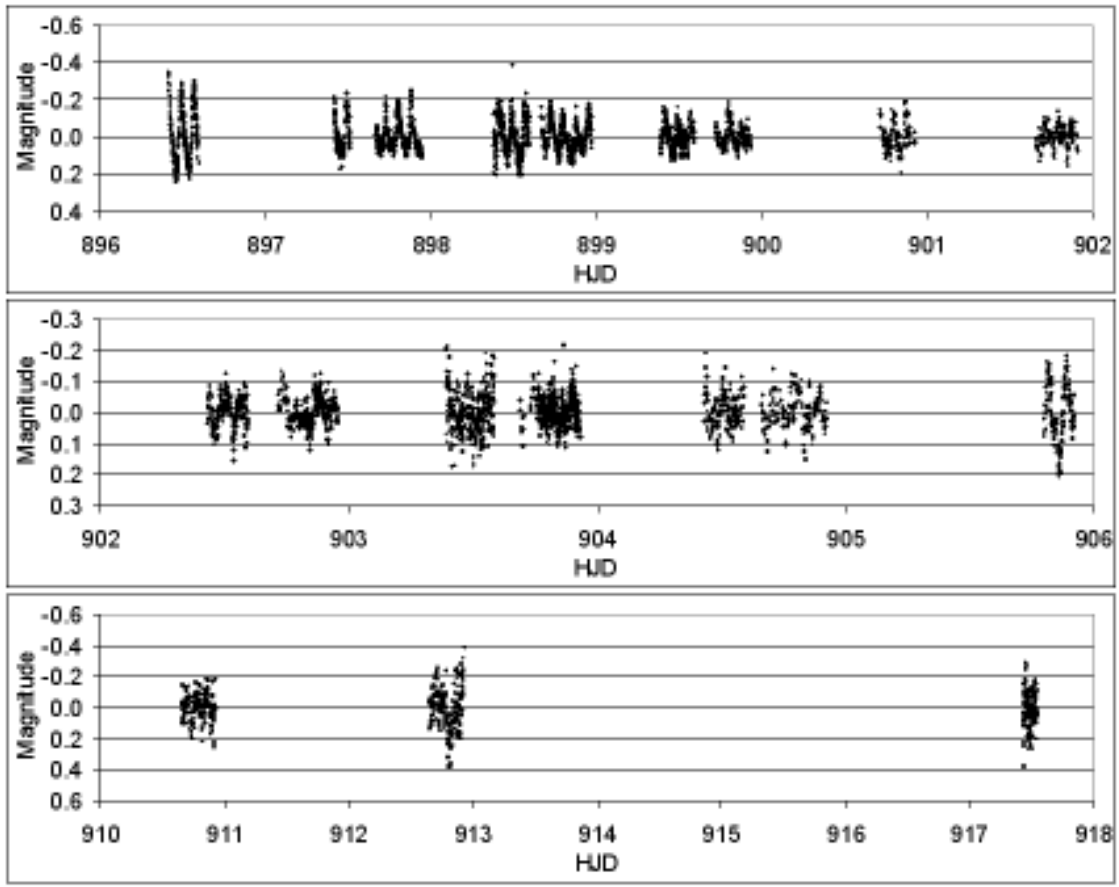

Figure 5: Light curves of all runs longer than 0.08 days after subtraction of mean and linear trends - (upper) JD 896-901, (middle) JD 902-905, (lower) JD 910-917.

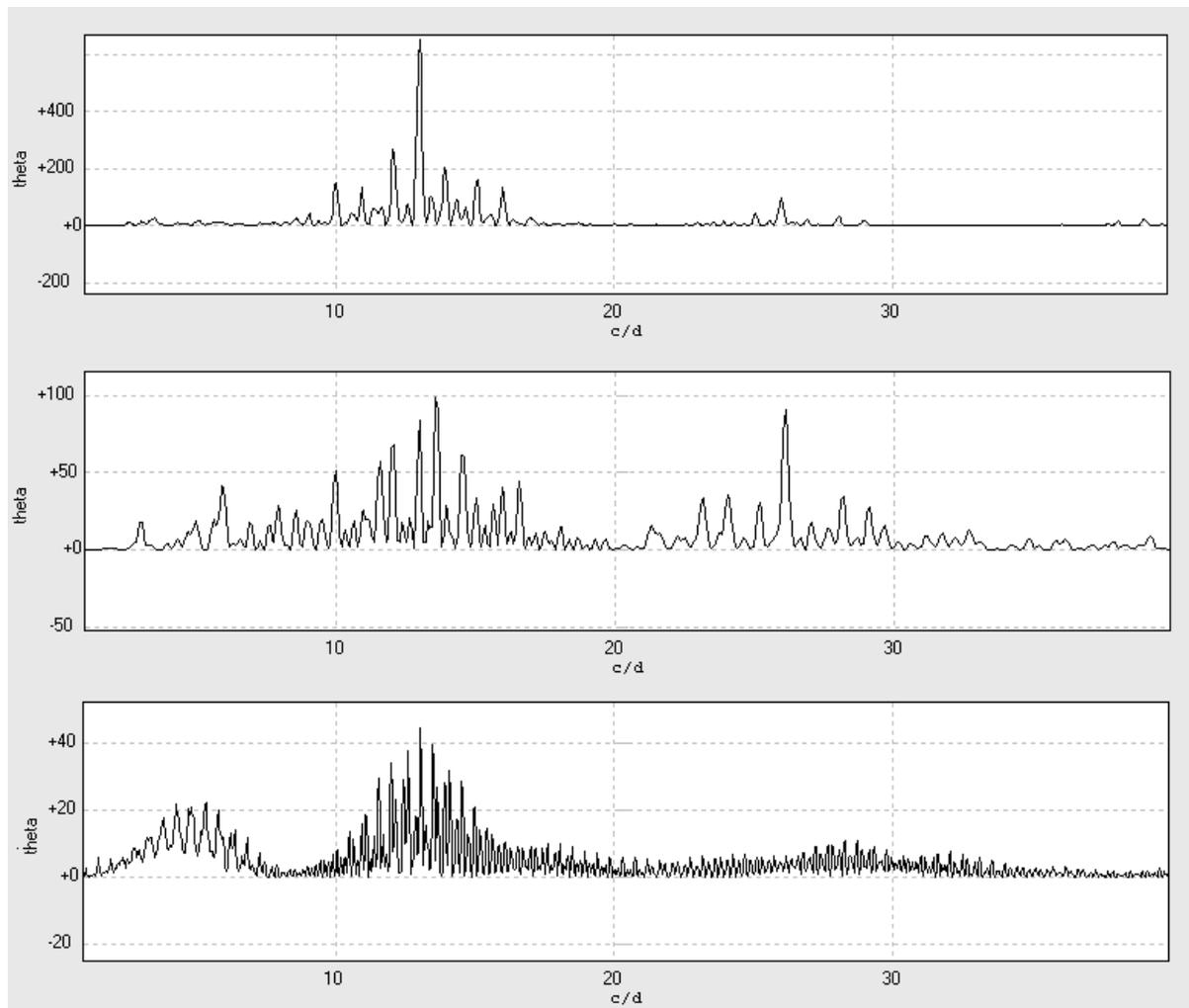

Figure 6: Power spectra - (upper) JD 896-901, (middle) JD 902-905, (lower) JD 910-917.

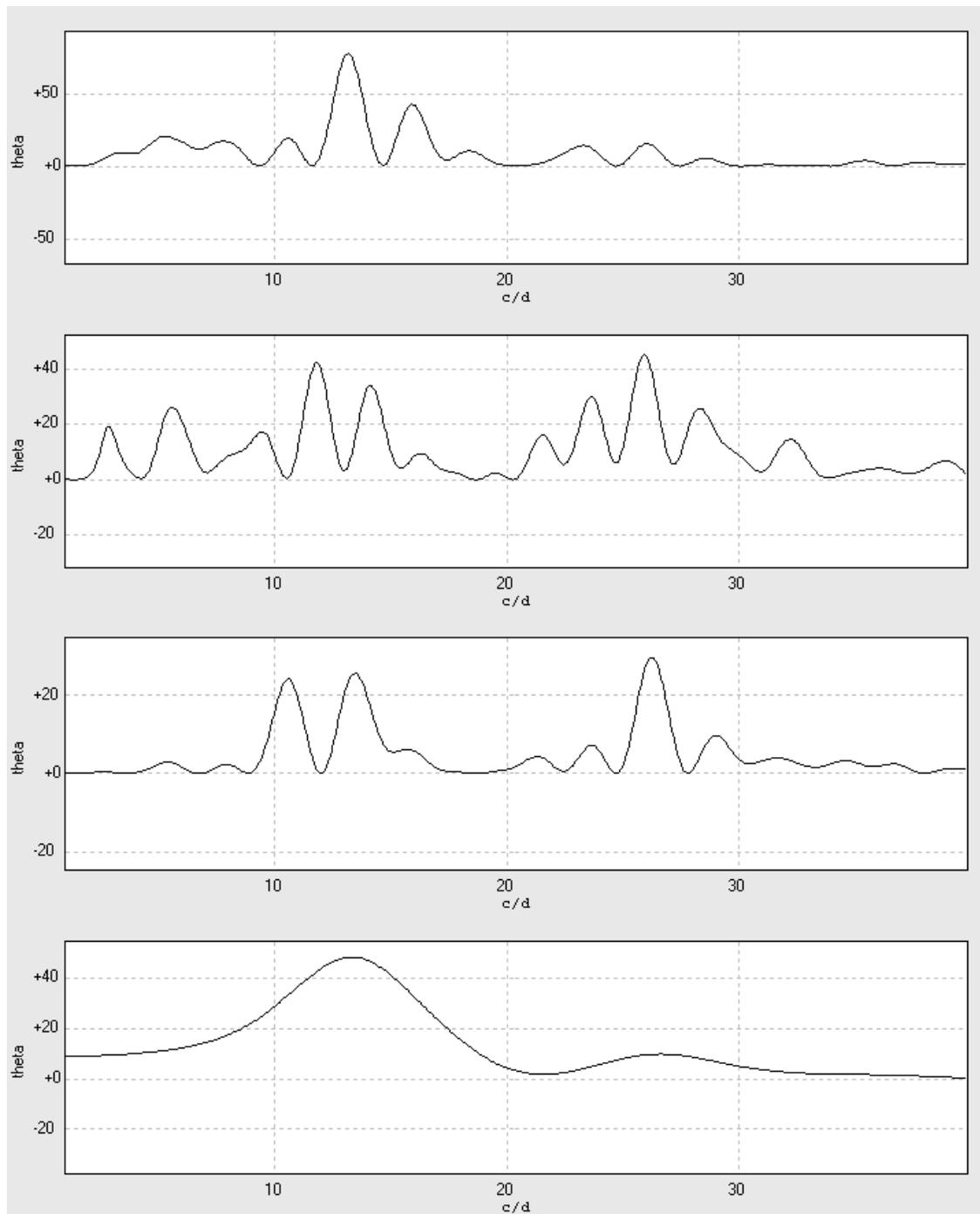

Figure 7: Power spectra - (upper) JD 902, (upper middle) JD 903, (lower middle) JD 904, (lower) JD 905.

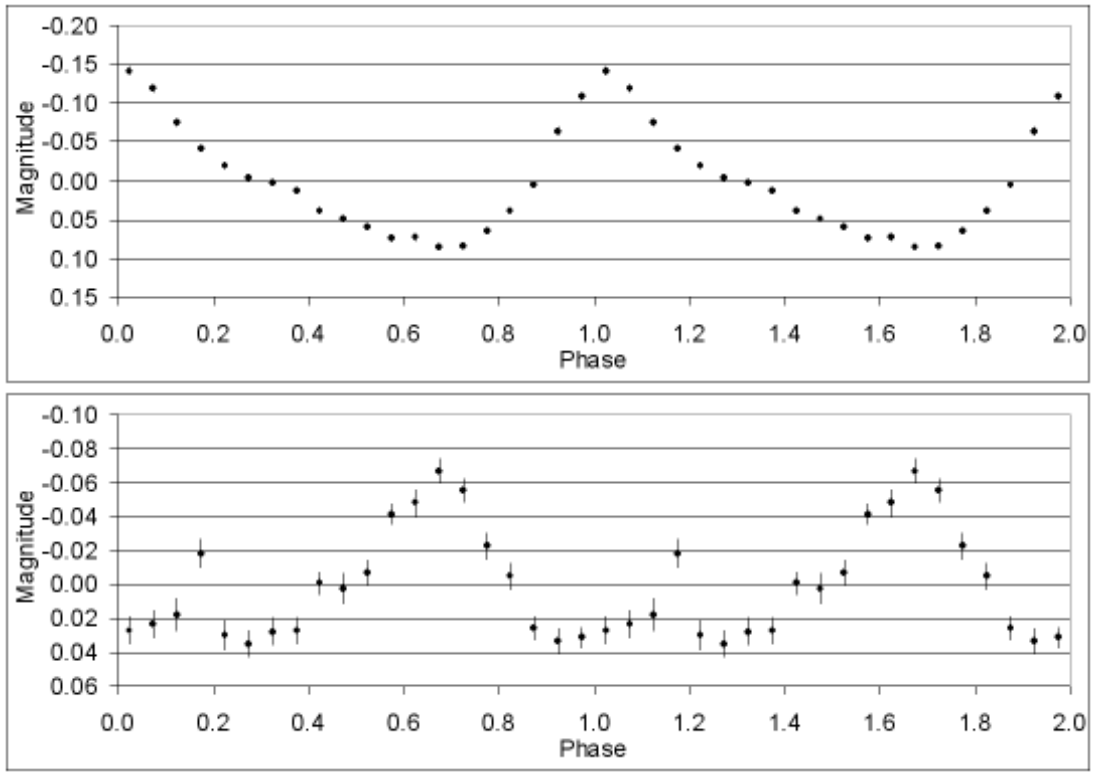

Figure 8: Phase diagrams - (upper) common superhumps averaged over the interval JD 896-901 showing 2 cycles, standard errors per bin are within the data points, (lower) late superhumps averaged over JD 903-917 with standard errors per bin.

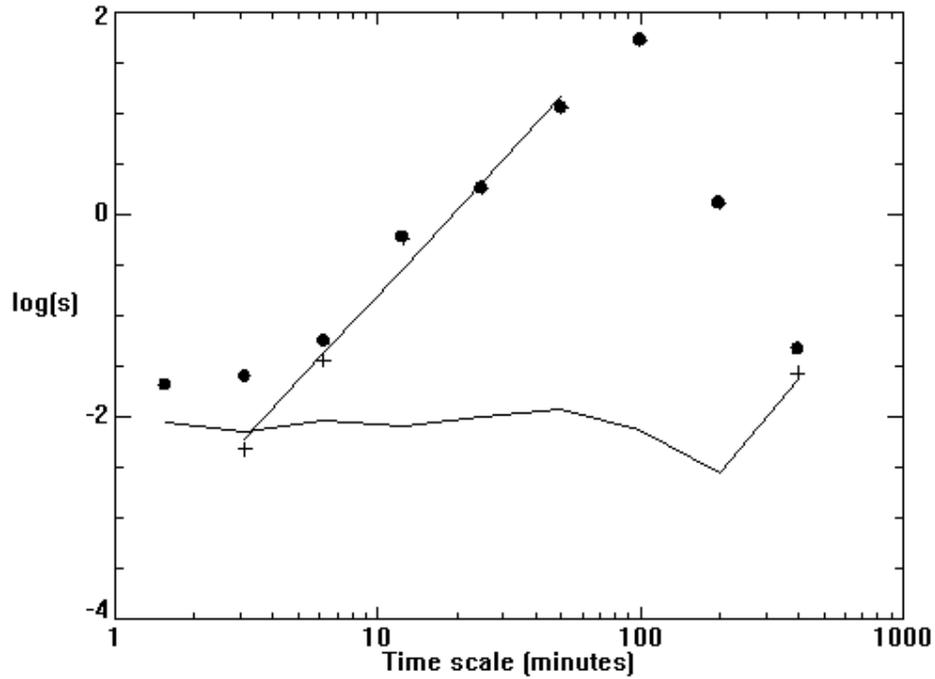

Figure 9: Scalegram of V1316 Cyg on JD 896 with the variable at magnitude 15.2 (upper, filled circles) and the equivalent scalegram of a 15.1 magnitude comparison star (lower track). The noise-corrected values are shown as crosses and are significantly different for only the weakest values of *log(s)*. The fit to the scalegram up to 60 minutes is shown by the straight line and this is used to determine the values of α and Σ. A similar process is used on all the other scalegrams.

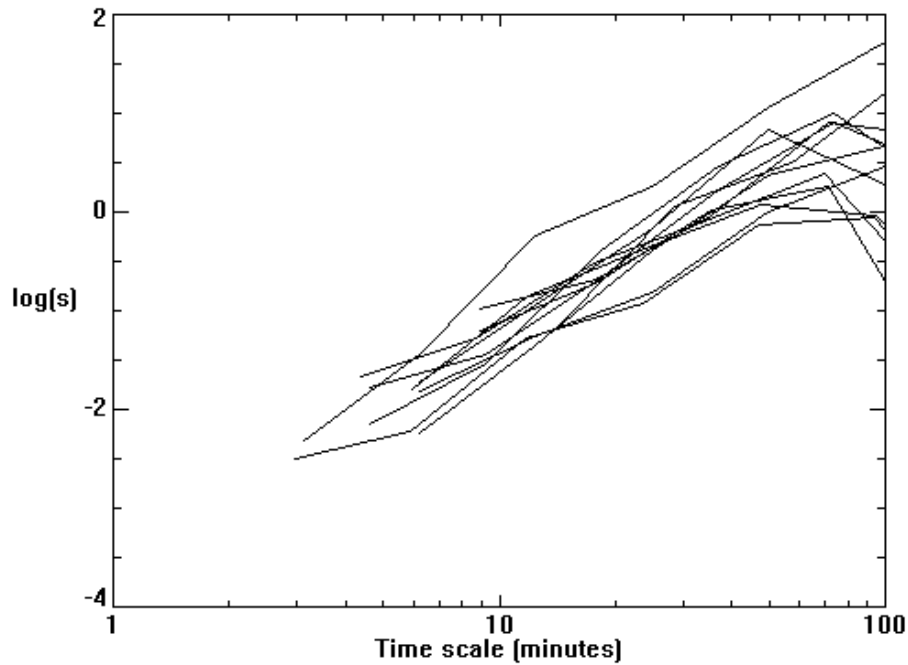

Figure 10: The scalegrams of all the data sets that provided reliable values of α and Σ.

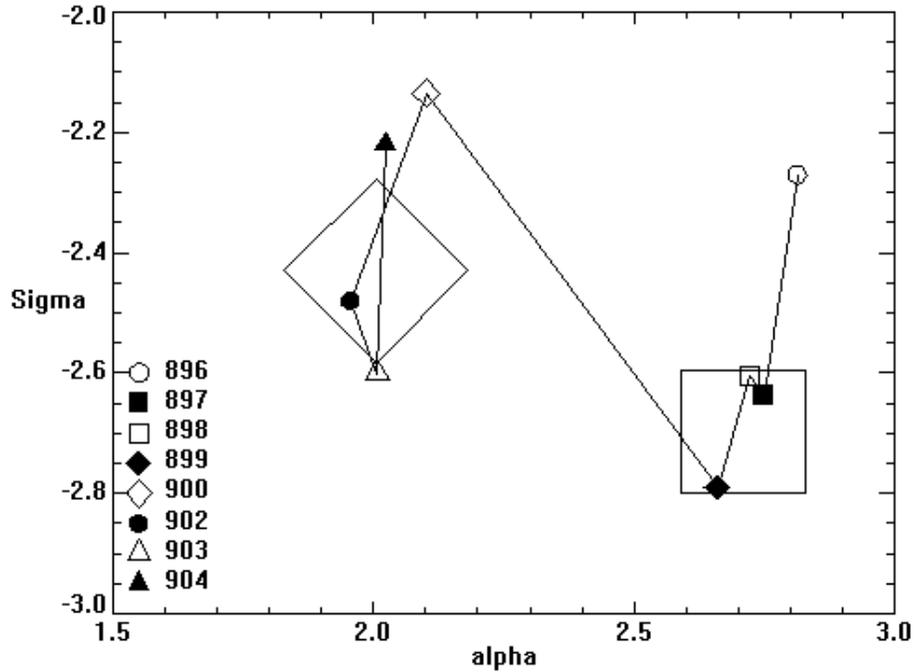

Figure 11: Evolution of the superoutburst in the α-Σ plane showing the weighted mean values from each day. The line connects the points in chronological order starting with the point on the top-right showing the data from Figure 9. The large square and diamond show the mean positions of the early (JD 896-899) and later (JD 900-904) data.

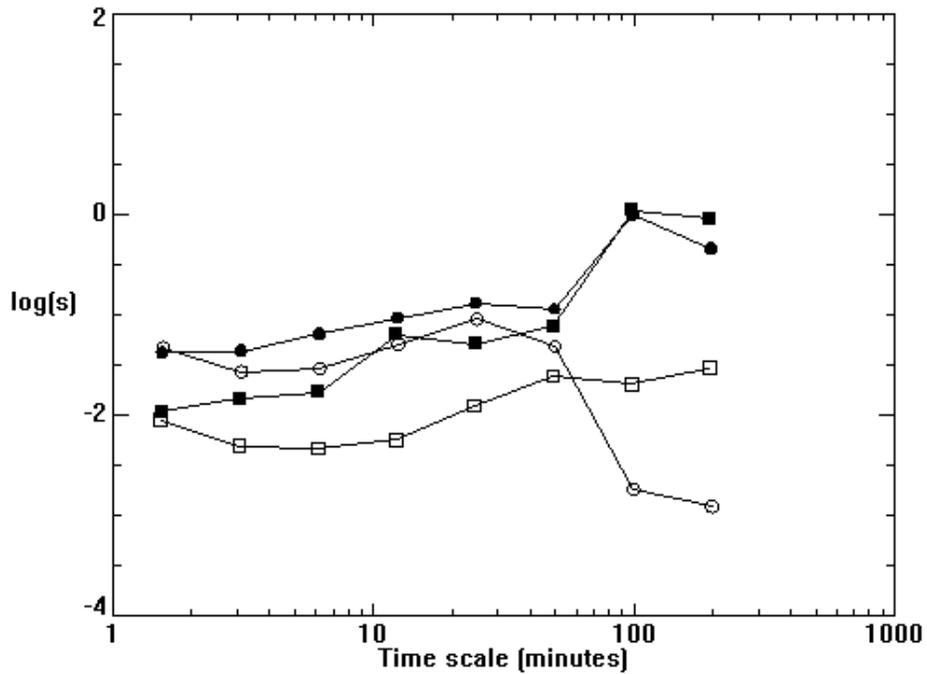

Figure 12: Scalegrams of the data from two brief outbursts, the first on JD 619 at magnitude 16.0 that occurred a year before the superoutburst (squares) and the second on JD 1003 at magnitude 16.1 that occurred shortly after (circles). The scalegrams of two comparison stars of similar magnitudes are shown as open symbols.